\documentclass[a4paper]{jpconf}
\usepackage{graphicx}
\usepackage{url}
\usepackage{hyperref}
\begin{document}
\title{Machine learning challenges in theoretical HEP}

\author{Stefano Carrazza}

\address{Theoretical Physics Department, CERN, Geneva, Switzerland}

\ead{stefano.carrazza@cern.ch\footnote{Preprint: CERN-TH-2017-212}}

\begin{abstract}
 In these proceedings we perform a brief review of machine learning
 (ML) applications in theoretical High Energy Physics (HEP-TH). We
 start the discussion by defining and then classifying machine
 learning tasks in theoretical HEP. We then discuss some of the most
 popular and recent published approaches with focus on a relevant case
 study topic: the determination of parton distribution functions
 (PDFs) and related tools. Finally, we provide an outlook about future
 applications and developments due to the synergy between ML and
 HEP-TH.
\end{abstract}

\section{Introduction}

Over the past several years machine learning (ML) has become one of
the most popular and powerful sets of techniques and tools used for
multidisciplinary scientific research. Such popularity has been
continuously growing in the past years thanks to the increasing number
of methodological developments, the availability of faster hardware
with strong computational capabilities such as modern GPUs and
coprocessors, and finally, the great interest and investment from the
private sector.

The recent enthusiasm has led to a new underlying code development
strategy where several tools based on ML are available as open source
projects, some examples are
TensorFlow~\cite{tensorflow2015-whitepaper},
scikit-learn~\cite{scikit-learn}, Keras~\cite{chollet2015},
Theano~\cite{theano} among others. Easy access to these tools has
simplified the integration of new modern techniques in many research
fields, in particular for those where a large amount of data is
available.

The dissemination of the innumerable applications and developments
based on ML has given way to conferences such as
ICML\footnote{\url{http://icml.cc}},
NIPS\footnote{\url{http://nips.cc}} and ACAT, but also to the
composition of specialized working groups, e.g the IML LHC working
group at CERN\footnote{\url{https://iml.web.cern.ch/}} which promotes
the integration of new techniques for specific requirements in
experimental analysis.

In the next sections we focus the discussion on ML in theoretical High
Energy Physics (HEP-TH). We start by defining the categories of
applications observed in that field. Then we take as a case study the
recent development achieved for the determination of parton
distribution functions (PDFs), where several techniques from ML are
employed. Finally, we conclude by listing the most plausible
development directions for the integration of ML in HEP-TH.

\section{Identifying machine learning applications in HEP-TH}

While classifying ML applications in experimental physics may seem
simple because the experimental analysis usually requires model
regression, classification and noise filtering to extract useful
information from large datasets. It becomes challenging when trying to
determine applications from the point of view of theoretical physics.

When talking about ML techniques in theoretical High Energy Physics
(HEP-TH) we may naively imagine that the usage of such tools is not
ideal. In fact, we can argue that the theoretical physicists are
trained to decode nature by constructing conjectures and building
theoretical frameworks which remap the complexity of the measured
observable in conceptual model rules. Therefore, this approach is in
contrast with a model determination through a ML black box model, of
which a physical interpretation is very difficult or nearly impossible
to achieve.

\begin{figure}
  \center
  \includegraphics[scale=0.6]{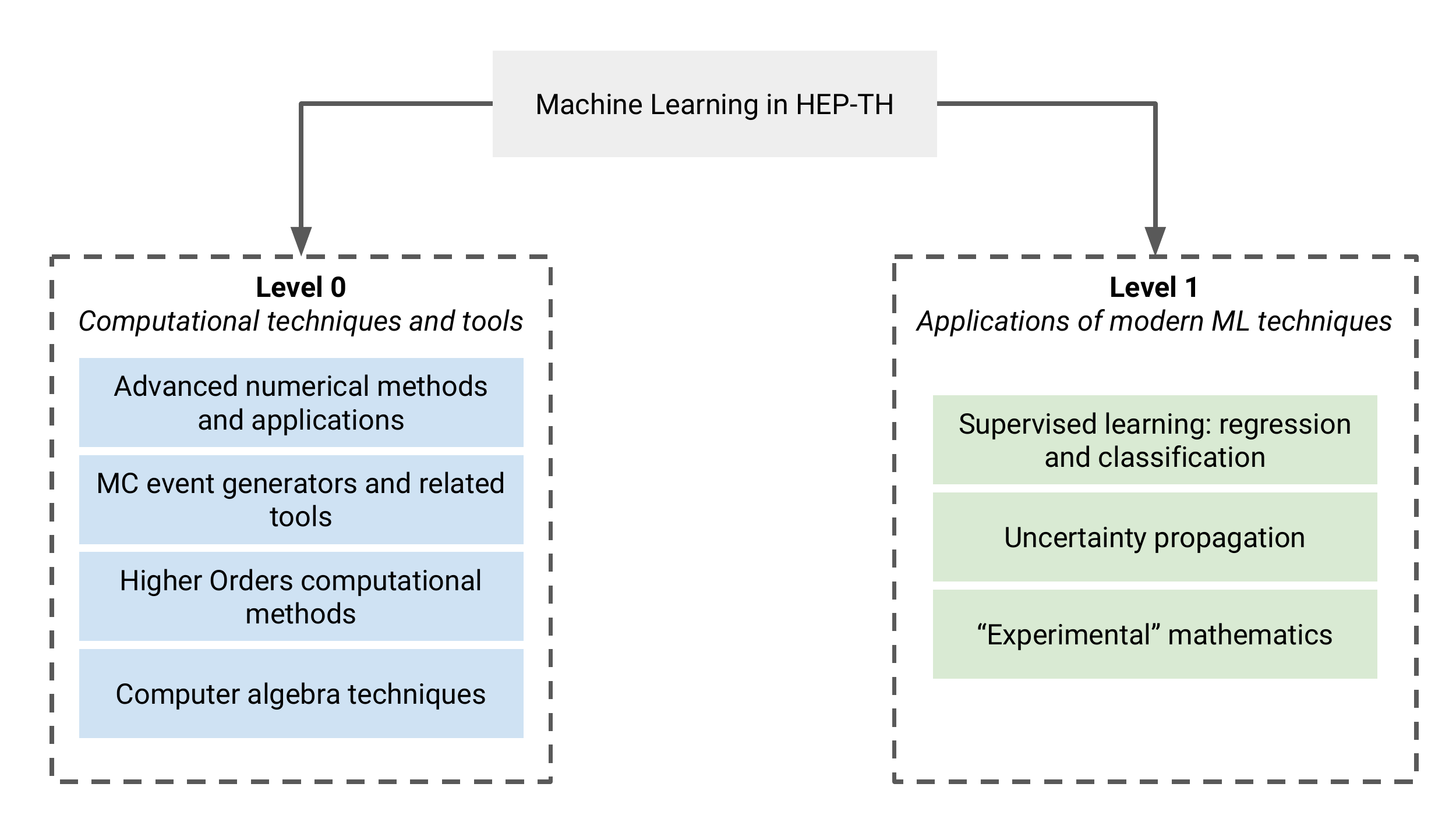}
  \caption{One possible graphical schematic representation of how
    machine learning is identified in HEP-TH applications.}
  \label{fig:mlscheme}
\end{figure}

However, nowadays such misleading interpretation is disappearing and
machine learning is becoming a set of useful tools that helps
achieving practical results in situations where the theory has a
strong computational requirement or when it requires the determination
of free parameters. So, before continuing our discussion, we think it
is important to define and classify how ML is translated in HEP-TH. In
Figure~\ref{fig:mlscheme} we summarize one plausible graphical
representation of the two branches of applications that we will
discuss now in turn: {\it Level-0} and {\it Level-1}.

%%%%%%%%%%%%%%%%%%%%%%%%%%%%%%%%
%%%%%%%%%%%%%%%%%%%%%%%%%%%%%%%%

\subsection{Level-0: computational techniques and tools}

The first level contains machine learning in terms of computational
techniques and tools. In this category we find the great majority of
topics presented in the Track 3 section of the ACAT conference. These
applications contain the most robust implications of computing in
theoretical physics. Due to the specific nature of the problems
addressed in HEP-TH these tools should be considered as part of ML
applications because their development has contributed in exhaustive
manner to the development of new techniques and methods in this field.

A non-exhaustive list of topics involved in {\it Level-0} are
summarized in the points below:

\begin{itemize}
\item Advanced numerical methods and applications: some examples
  are algorithms for Monte Carlo and Quasi Monte Carlo integration,
  techniques for subtraction schemes and regularization of Feynman
  integrals, e.g.~\cite{Boughezal:2016zws}, resummation
  techniques~\cite{Altarelli:2008aj}.
\item Monte Carlo event generators: in this category we have the
  methodological developments established by MC codes such as
  POWHEG~\cite{Nason:2004rx}, MadGraph\_aMG5~\cite{Alwall:2014hca},
  Pythia~\cite{Sjostrand:2006za}, Herwig~\cite{Bahr:2008pv},
  MCFM~\cite{Boughezal:2016wmq} and several others. The possibility to
  reuse events independently from the parton distribution functions
  thanks to reweighting and weight storage techniques such as
  APPLgrid~\cite{Carli:2010rw} and FastNLO~\cite{Kluge:2006xs}.
\item Higher order computational methods: numerical techniques for
  $N$-loop integrals as OneLoop~\cite{vanHameren:2010cp},
  QCDLoop~\cite{Carrazza:2016gav}, LoopTools~\cite{Hahn:1998yk};
  parton level generators at NNLO, e.g.~DYNNLO~\cite{Catani:2009sm}
  and the more recent N3LO~\cite{Anastasiou:2015ema}.
\item Computer algebra techniques: some examples of algebra systems
  developed for the HEP-TH community as FORM~\cite{Ruijl:2017dtg} and
  QGRAF~\cite{Nogueira:1991ex} for the translation of Feynman diagram
  rules into compact analytic expressions for its numerical
  evaluation.
\end{itemize}

\subsection{Level-1: applications of ML modern techniques}

This second level contains the ML applications in ``sensu stricto''
i.e.~using ML modern techniques used in data sciences. This kind of
application usually requires hybrid projects where experimental data
and theory are included together. Nevertheless, during the last few
years there has been a strong development of very successful tools
based on these techniques.

Some examples for this category are:

\begin{itemize}
\item Supervised learning, such as regression and classification:
  parton distribution~\cite{Ball:2017nwa} and fragmentation
  functions~\cite{Bertone:2017tyb} determination, Monte Carlo
  tunes~\cite{Skands:2014pea}, reweighting techniques, jet
  discrimination through deep convolutional neural
  networks~\cite{Komiske:2016rsd}.
\item Techniques for uncertainty estimation and combination: in this
  category we find several tools from the PDF4LHC15
  recommendation~\cite{Butterworth:2015oua}, and some recent methods
  to provide a reliable MC uncertainty for simulations, e.g.~by
  modeling jet predictions~\cite{Carrazza:2017bjw}.
\item ``Experimental'' mathematics: applications based on machine
  learning optimization algorithms which may lead to the determination
  of multivariate densities~\cite{likas,Krefl:2017wgx} and sampling
  for integral evaluation~\cite{Bendavid:2017zhk}.
\end{itemize}

In the next section we describe the recent innovative achievements
obtained by the parton distribution function community in HEP-TH
using ML methods.

\section{Case study: the proton structure determination}

The Quantum Chromodynamics (QCD) theory describes the proton structure
in terms of partons, e.g.~quarks and gluons, but due to the
non-perturbative regime of confinement we are unable to evaluate from
QCD the momentum fraction of the proton carried by each
parton. However, in order to avoid our lack of knowledge we introduce
the concept of parton distribution functions (PDFs). These PDFs are
then inferred from data of all relevant processes, together with the
theoretical knowledge on how they affect the cross section which can
be calculated approximately in perturbation theory.

\subsection{The NNPDF approach}

%%%%%%%%%%%%%%%%%%%%%%%%%%%%%%%%
\begin{figure}
  \center
  \includegraphics[scale=0.6]{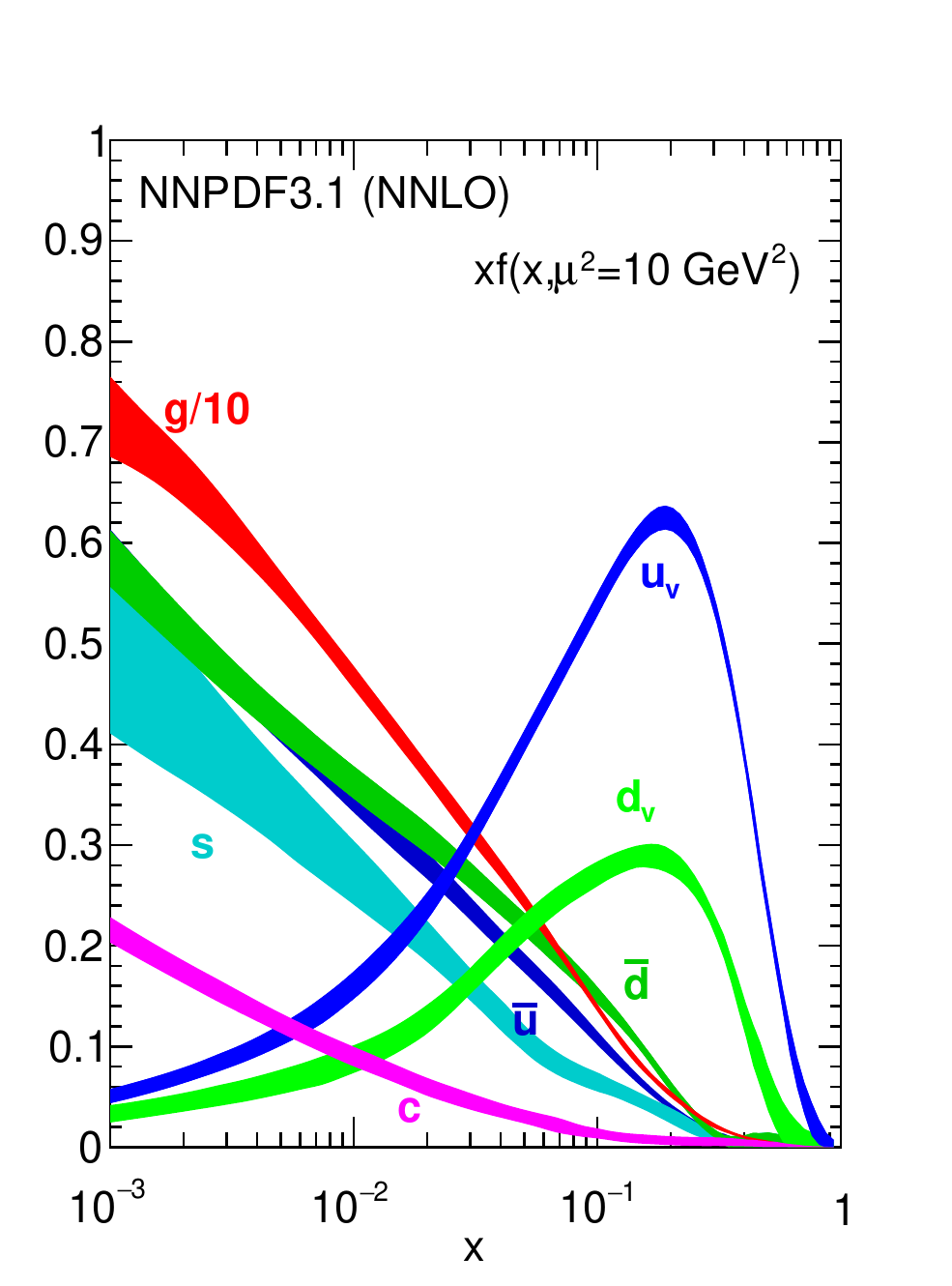}\includegraphics[scale=0.6]{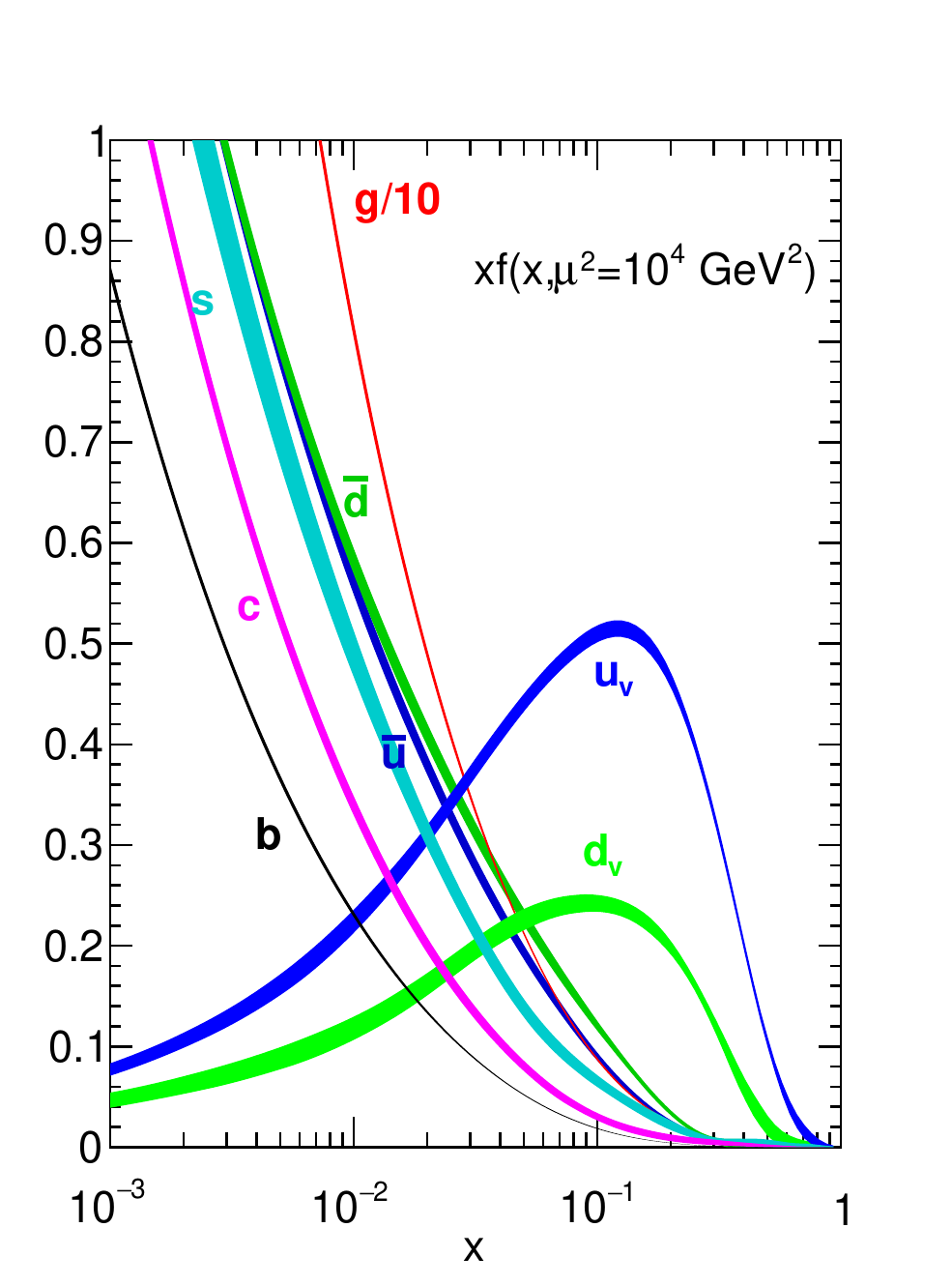}
  \caption{The NNPDF3.1 NNLO PDFs, evaluated at $\mu^2=10$ GeV$^2$
    (left) and $\mu^2=10^4$ GeV$^2$ (right) from~\cite{Ball:2017nwa}.}
  \label{fig:nnpdf31}
\end{figure}
%%%%%%%%%%%%%%%%%%%%%%%%%%%%%%%%

The NNPDF collaboration uses ML techniques to obtain a PDF
determination. In contrast to other problems in ML, where finding an
accurate and fast algorithm is enough, we are not only interested in
the best PDF fit, but also in obtaining an uncertainty estimate of the
PDF determination. The procedure we employ is described in full
details in~\cite{Ball:2014uwa} and can be summarized as:
\begin{itemize}

\item Monte Carlo generation of artificial data. Experimental data,
  with central values, errors and their correlations are used to
  generate further artificial data, consistent with the covariance
  matrix provided by the experiment.
  
\item Neural network fit to artificial data. A genetic algorithm is
  used to fit each artificial data set to a neural network
  representing the PDF (more details in~\cite{Carrazza:2017udv}).

\item Predictions are later obtained by computing statistical
  estimators (such as means, quantiles and standard deviations) over
  the set of neural networks. Figure~\ref{fig:nnpdf31} illustrates the
  current state of the art PDFs, NNPDF3.1~\cite{Ball:2017nwa},
  obtained with the NNPDF framework.
  
\end{itemize}

The points above summarize the main difficulties in PDF fits. The
first consists in the inclusion of multiple experimental data which
introduces indirect constraints on the PDFs. The data included in such
fits are based on several datasets measured during the past decades
and based on different physical processes: deep-inelastic scattering,
fixed target Drell-Yan and hardronic data. This data is obtained
through diverse experimental techniques and statistical analyses,
therefore yielding possible inconsistencies and tensions among
themselves. In order to limit these effects the NNPDF methodology
propagates the uncertainty of the experimental data on the PDFs by
performing the Monte Carlo artificial replica generation based on the
covariance matrix provided by each experiment.

The second most relevant problem consists in the choice of an unbiased
functional form able to adapt and allow the propagation of data
uncertainties into PDF errors while keeping under control physical
requirements such as momentum sum rule conservation and positivity
constraints. The NNPDF collaboration has successfully employed neural
networks based on feed-forward multilayer perceptron architecture to
model each single PDF flavor entering the fit procedure. The large
number of parameters from these PDFs are then trained through a
genetic optimization algorithm.

NNPDF has released several PDF sets of global unpolarized
determinations in the last
years\footnote{\url{http://nnpdf.mi.infn.it/}}, together with the more
recent polarized PDFs~\cite{Nocera:2014gqa} and Fragmentation
Functions~\cite{Bertone:2017tyb}. One of the current most important
tasks of the NNPDF collaboration is to improve the performance and
quality of the optimization having in mind that in the next years the
number of the new measurements from LHC will increase.

\subsection{PDF4LHC15 tools for LHC Run II}

Another issue concerning PDF determination consists in deciding which
set of PDF should be used for the calculation of a PDF-dependent
quantity. The question addressed here is how to obtain the best
combined PDF uncertainty from individual PDF sets. This kind of issue
is studied by the PDF4LHC working group which releases a
recommendation document usually every three years. The latest one
published in 2015~\cite{Butterworth:2015oua} aims to release combined
PDF sets based on methods with clear statistical interpretation.

The PDF4LHC15 prescription starts by constructing a prior Monte Carlo
combined set of PDFs derived from global determinations, namely
MMHT14~\cite{Harland-Lang:2014zoa}, CT14~\cite{Dulat:2015mca} and
NNPDF3.0~\cite{Ball:2014uwa}. These sets satisfy requirements such as:
similar dataset, DGLAP solution and $\alpha_s$. The prior construction
is obtained by converting the Hessian sets to MC replicas following
the procedure described in~\cite{Watt:2012tq}. After that we apply
algorithms to reduce the redundant information stored in the prior and
deliver sets of PDFs with the same properties of the prior but smaller
number of replicas.

%%%%%%%%%%%%%%%%%%%%%%%%%%%%%%%%
\begin{figure}
  \center \includegraphics[scale=0.6]{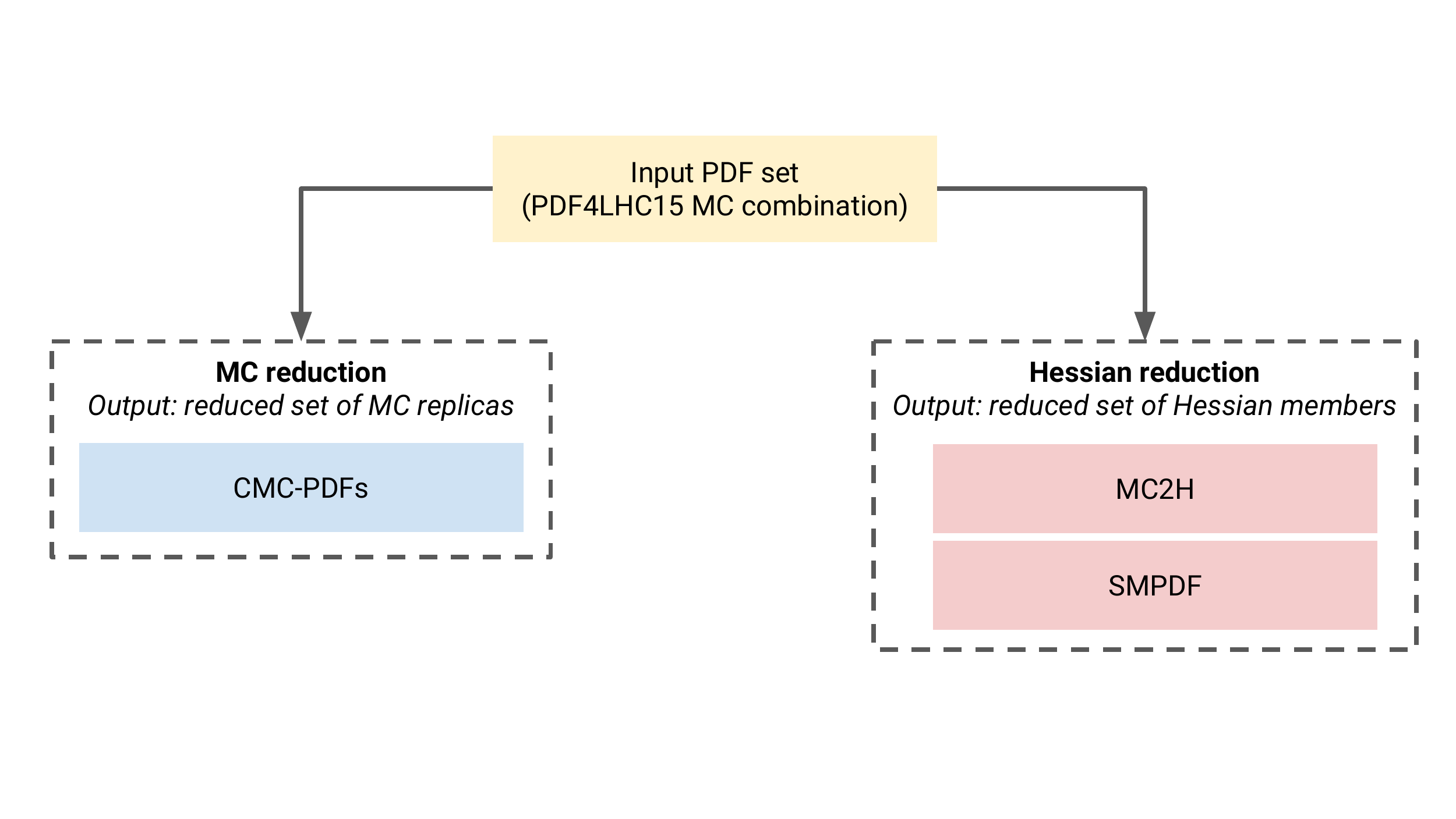}
  \vspace{-1cm}
  \caption{Graphical representation of the reduction algorithms used
    in the PDF4LHC15 recommendation~\cite{Butterworth:2015oua}.}
  \label{fig:pdf4lhc}
\end{figure}
%%%%%%%%%%%%%%%%%%%%%%%%%%%%%%%%

\subsection*{Compression algorithm}

The CMC-PDFs~\cite{Carrazza:2015hva} implement the compression
algorithm of MC replicas. This algorithm is designed to extract a
subset of replicas which preserves as much as possible the underlying
statistical distribution of the prior MC PDF set. This algorithm
pre-computes for the input PDF the moments (central value, variance,
skewness and kurtosis), the statistical distance (Kolmogorov distance)
and the correlations for each flavor in a grid of $x$ points. These
estimators are then compared to subsets of replicas selected by a
genetic algorithm. The procedure terminates when a tolerance value for
the error function comparing these estimators is flat. From a
practical point of view in the context of the PDF4LHC15 the initial
$N_{\rm rep}=900$ replicas of the prior set have been reduced to 100
replicas. Further technical investigations based on clustering
algorithms have also been performed in~\cite{Carrazza:2016sgh}.

\subsection*{MC2H}

The MC2H algorithm~\cite{Carrazza:2015aoa} was first introduced to
convert MC PDF sets into a Hessian representation. This algorithm
performs the principal component analysis (PCA) of the PDF covariance
matrix in a predefined grid of $x$ nodes for all flavors at a given
initial scale. The eigenvectors obtained from the PCA are the basis of
the Hessian representation. This representation consists in simple
linear combinations of the input MC replicas. The MC2H procedure also
allows the reduction of the number of required replicas of the input
PDF set because we can reject the eigenvectors associated to small
eigenvalues therefore obtaining a reduced set of Hessian replicas.

\subsection*{SMPDF}

Starting from the MC2H algorithm we have developed the Specialized
Minimal PDFs (SMPDF)~\cite{Carrazza:2016htc} which consists in
obtaining the smallest set of Hessian replicas for a given physical
process. The SMPDF algorithm performs an interactive PCA reduction on
top of the PDF covariance matrix computed in a grid of $x$ points
determined as the region where the PDF-process correlation is
maximal. We have also provided a public web-interface for the
construction of SMPDFs available and documented in
~\cite{Carrazza:2016wte}.

\section{Outlook}

The advantages and results obtained thanks to ML have been essential
for several developments of this research field from PDF determination
to Monte Carlo event generation. In the next months and years new
applications will be achieved. It is already possible to summarize the
two main directions for these developments.

The first consist in the development of tools for the estimation and
propagation of uncertainties. New methods to deal with uncertainty
will possibly improve the determination of PDFs among other
applications. This point also includes the new ideas about reweighting
techniques in the context of higher order calculations.

The second development branch consists in the construction of new
modern neural network architectures for problem specific applications
together with efficient new gradient based methods. Such tools will
open the possibility to obtain better methods for multidimensional
density estimation. Consequently, we will obtain better sampling
algorithms relevant for improved and faster integration algorithms in
Monte Carlo event generators and similar tools.

In these proceedings we provided a short overview of successful
applications of ML in HEP-TH but this is just the prelude of a new era
where ML and HEP-TH are in synergy producing innovative and unique
results.

\section*{Acknowledgements}

S.~C. is supported by the HICCUP ERC Consolidator grant (614577) and
by the European Research Council under the European Union's Horizon
2020 research and innovation programme (grant agreement n$^{\circ}$
740006).

\end{document}